\documentclass[twocolumn,preprintnumbers,amsmath,amssymb,superscriptaddress,prl]{revtex4}
\usepackage[english]{babel}
\usepackage[dvips]{graphicx}
\usepackage{dcolumn}
\usepackage{bm}
\usepackage[dvips]{epsfig}
\usepackage{hyperref}
\usepackage[normalem]{ulem}
\usepackage{color,amssymb}

\begin{document}

\title{Criterion for fingering instabilities in colloidal gels}

\author{Thibaut Divoux}
   \affiliation{MultiScale Material Science for Energy and Environment, UMI 3466, CNRS--MIT, 77 Massachusetts Avenue, Cambridge, Massachusetts 02139, USA} 
  \affiliation{Department of Civil and Environmental Engineering, Massachusetts Institute of Technology, Cambridge, MA 02139}
\author{Asheesh Shukla}
  \affiliation{MultiScale Material Science for Energy and Environment, UMI 3466, CNRS--MIT, 77 Massachusetts Avenue, Cambridge, Massachusetts 02139, USA}
\author{Badis Marsit}
  \affiliation{Department of Mechanical Engineering, Massachusetts Institute of Technology, Cambridge, MA 02139}
\author{Yacouba Kaloga}
  \affiliation{MultiScale Material Science for Energy and Environment, UMI 3466, CNRS--MIT, 77 Massachusetts Avenue, Cambridge, Massachusetts 02139, USA}
\author{Irmgard Bischofberger}
  \affiliation{Department of Mechanical Engineering, Massachusetts Institute of Technology, Cambridge, MA 02139}

\date{\today}

\begin{abstract}
We sandwich a colloidal gel between two parallel plates and induce a radial flow by lifting the upper plate at a constant velocity. Two distinct scenarios result from such a tensile test: ($i$)~stable flows during which the gel undergoes a tensile deformation without yielding, and ($ii$)~unstable flows characterized by the radial growth of air fingers into the gel. We show that the unstable regime occurs beyond a critical energy input, independent of the gel's macroscopic yield stress. This implies a local fluidization of the gel at the tip of the growing fingers and results in the most unstable wavelength of the patterns exhibiting the characteristic scalings of the classical viscous fingering instability. Our work provides a quantitative criterion for the onset of fingering in colloidal gels based on a local shear-induced yielding, in agreement with the delayed failure framework.
\end{abstract}

\maketitle

The displacement of a more viscous fluid  by a less viscous one in a confined geometry can induce the formation of complex patterns \cite{Saffman:1958,Paterson:1981,Chen:1987}. Even in the simplest case of two Newtonian fluids, this viscous fingering instability keeps yielding new discoveries \cite{Pihler:2012,Bischofberger:2014,Bihi:2016,Rabbani:2018,Alert:2019,Escala:2019,Videbaek:2019}. 
When the displaced fluid is replaced with a non-Newtonian one, the finger growth dynamics can be significantly altered \cite{Bohr:1994,McCloud:1995,Kondic:1998,Lindner:2002,Puff:2002,Shaukat:2009,Abdelhaye:2012,VanDamme:2002}. 
For instance, in the displacement of a viscoelastic fluid by a Newtonian one, a transition from viscous fingering to fracture occurs above a critical Deborah number, driven by the release of elastic stresses \cite{Lemaire:1989,Lemaire:1991,Zhao:1993,Mora:2010,Foyart:2013,Ligoure:2013}. 
While a consensus has been reached for the mechanism responsible for this transition in viscoelastic fluids with a finite relaxation timescale, we are far from a comprehensive description of fingering instabilities in another class of complex materials, yield stress fluids, which exhibit a solid-like behavior at rest and a solid-to-liquid transition beyond a critical stress $\sigma_c$ \cite{Coussot:2014,Bonn:2017,Coussot:2018}.
In dense foams and emulsions displaced by air, for example, the pattern morphology associated with fingering instabilities is strongly rate-dependent \cite{Park:1994,Lindner:2000}. In tensile tests of dense microgels, where the samples are sandwiched between two parallel plates that get separated at a constant velocity, the yield stress can suppress the instability \cite{Barral:2010}. The transition between stable and unstable displacement currently lacks a theoretical explanation  \cite{Coussot:1999b,Barral:2010}. Moreover, the most unstable wavelength of the pattern has been alternatively reported as being set by the yield stress, or being independent of the yield stress \cite{Lindner:2000,Derks:2003,Maleki:2005}, which calls for further experimental investigations.

In this Letter, we report a comprehensive description of both the criterion for stabilization and the characteristics of the most unstable wavelength for a colloidal gel displaced by air. We show that the onset of the viscous fingering instability occurs at a critical energy input that is independent of the gel's yield stress, implying that the mechanism inducing unstable growth is determined locally at the particle scale. Remarkably, the most unstable wavelength $\lambda_{\rm c}$ obeys the scaling of a Newtonian fluid, which indicates a complete fluidization of the gel at the locus of finger growth. This observation, in agreement with a local yielding criterion, is further confirmed by the power-law scaling of the normal force relaxation occurring during the tensile test. Our results provide a comprehensive framework for fingering instabilities in colloidal gels based on a mechanism that is specific to attractive systems, which suggests that there will not be a universal framework accounting for viscous fingering instabilities in all yield stress fluids.

\begin{figure*}[!t]
\centering
\includegraphics[width=0.8\linewidth]{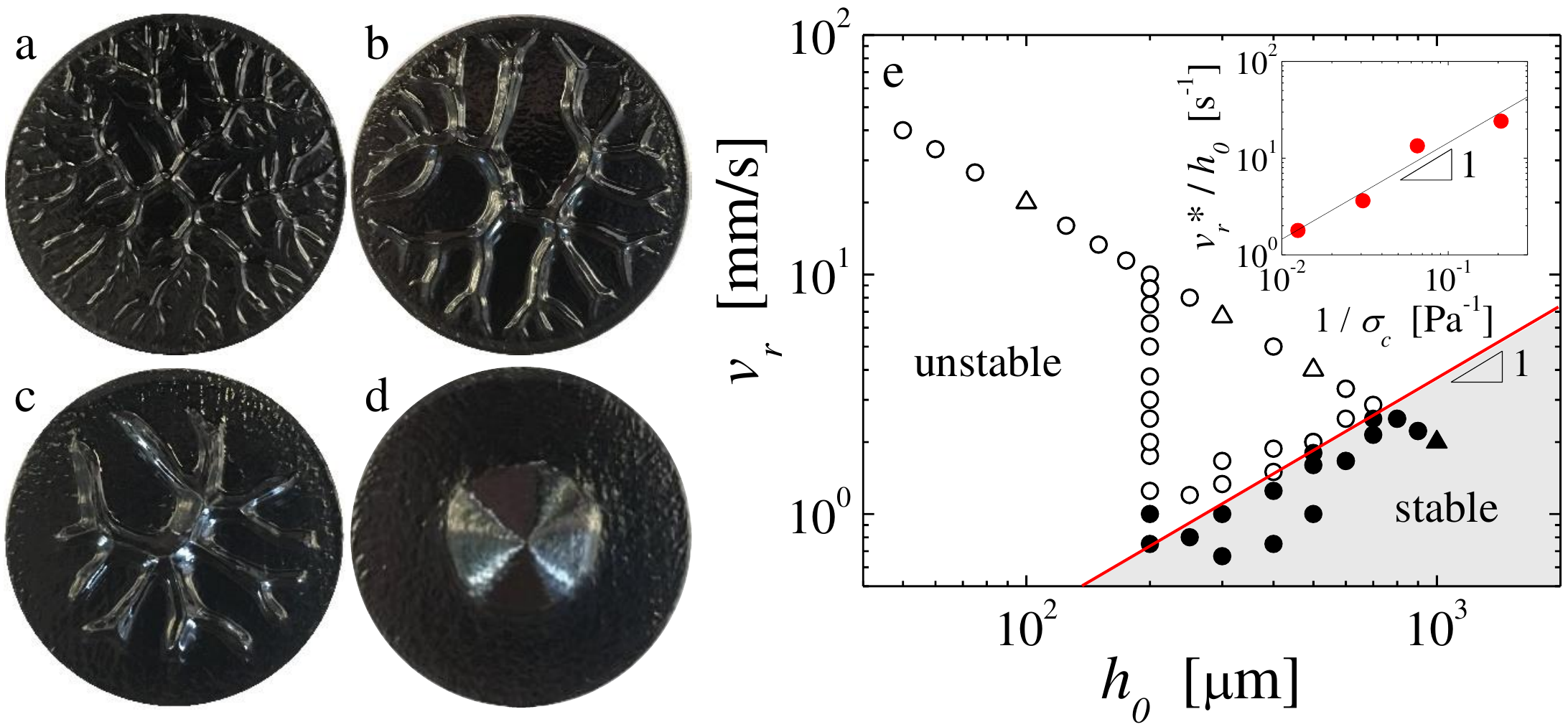}
\caption{(a)--(d) Patterns obtained at a lift velocity $v_l=200~\mu$m/s for increasing initial gap thicknesses $h_0$= 100, 300, 500 and 900~$\mu$m in tensile tests performed with a 8\% wt.~carbon black gel placed between two parallel plates of radius $R=20$~mm. (e) Stability diagram: radial velocity $v_r$ versus initial gap thickness $h_0$. Closed symbols denote stable conical patterns, open symbols denote unstable fingering patterns. The data points corresponding to the four images are indicated as triangles. The red line denotes the critical velocity $v_r^*$ separating the stable from the unstable regime. Inset: $v_r^*/h_0$ versus the inverse of the yield stress $\sigma_c^{-1}$ for carbon black gels at 4\%, 6\%, 8\% and 10\% wt. $\sigma_c$ is determined as the crossover of $G'$ and $G''$ during a stress sweep from 1~Pa to 100~Pa (frequency $f=1$~Hz, waiting time of 5~s per point).  
\label{fig1}}
\end{figure*} 
The colloidal gel consists of carbon black particles (Vulcan XC72R, Cabot) of typical diameter $a=0.5$~$\mu$m suspended at weight fractions ranging from 4\% to 10\% wt.~in light mineral oil (Sigma Aldrich; viscosity $\eta_s=20$~mPa.s, density $\rho_s=0.838$~g/ml). 
Due to attractive van der Waals forces \cite{Waarden:1950,Hartley:1985}, carbon black particles form a space-spanning gel network of fractal dimension $d_f\simeq 2.2$, whose linear viscoelastic properties are characterized by a frequency-independent elastic response \cite{Trappe:2000,Prasad:2003,Gibaud:2010,Grenard:2014}. Under external stresses lower than the yield stress $\sigma_c$, the gel behaves as a solid. For $\sigma> \sigma_c$, the gel flows. This shear-induced yielding transition is time-dependent, spatially heterogeneous and characterized by activated dynamics \cite{Gibaud:2010,Sprakel:2011,Grenard:2014,Perge:2014b,Gibaud:2016}. Here, we place a carbon black gel between a stainless steel parallel-plate geometry of radius $R=20$~mm, whose upper plate is connected to a stress-controlled rheometer (DHR-3, TA Instruments). The temperature of the lower plate is fixed to 25$^{\circ}$C by Peltier elements. To account for the sensitivity of carbon black gels on shear history \cite{Ovarlez:2013,Helal:2016,Narayanan:2017,Hipp:2019}, the sample is fully fluidized prior to each test by applying a large shear stress $\sigma=100$~Pa for 30~s. The stress is subsequently swept down to $\sigma=0$~Pa at a rate of 1~Pa/s, while the gel reforms. This protocol yields a reproducible initial state \cite{Helal:2016}. The resulting viscoelastic moduli $G'_0$ and $G''_0$ of the gel are determined during 1~min by small amplitude oscillations (strain amplitude $\gamma=0.1$\%, frequency $f=1$~Hz) [Fig.~S1 in Supplemental Material]. We then perform a tensile test at a constant lift velocity $v_l$ during which we record the normal force $F_{N}$. The plate separation yields a roughly symmetric pattern on both plates, which is photographed.

\begin{figure}
\centering
\includegraphics[width=0.9\linewidth]{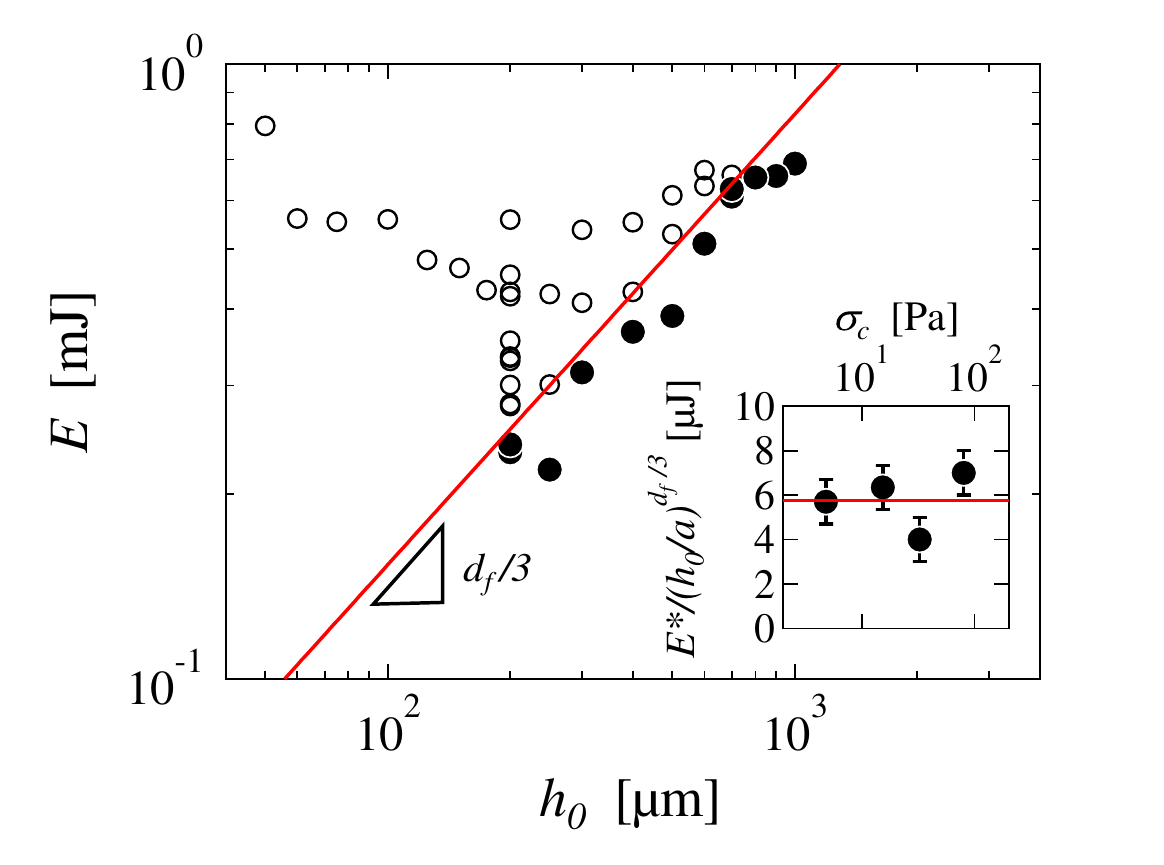}
\caption{Energy input $E$ associated with the separation of two plates sandwiching a 8\% carbon black gel initially separated by $h_0$. Closed symbols denote stable patterns, open symbols denote unstable patterns. The red line defines the critical energy $E^*$ beyond which fingers grow. $E^*$ exhibits a power law with $h_0$, with an exponent $d_f/3$, where $d_f=2.1$. Inset: critical energy per particle $E^*/(h_0/a)^{d_f/3}$ versus yield stress for carbon black gels at 4\%, 6\%, 8\% and 10\% wt.
\label{fig2}}
\end{figure}

Two main types of patterns occur in the tensile tests for a 8\% wt.~carbon black gel, depending on the lift velocity $v_l$ and the initial gap thickness $h_0$: ($i$)~unstable patterns characterized by highly branched structures [Fig.~\ref{fig1}(a)-(c)], and ($ii$) stable patterns characterized by conical piles of carbon black gel at the center of the plates [Fig.~\ref{fig1}(d)]. Unstable patterns result from instabilities at the air/gel interface at the periphery of the parallel plate geometry and are observed for high lift velocities and small initial gap thicknesses. Small perturbations here have a positive growth rate and grow into large-scale air fingers along the radial direction, leaving behind regions depleted in gel. Stable patterns result from a stable displacement of the gel under the tensile flow generated by the plate separation, where perturbations at the interface get damped out. They are observed for low lift velocities and large initial gap thicknesses. Such stable patterns do not occur in Newtonian fluids, but were reported in soft glassy materials such as dense microgels and mortar pastes \cite{Barral:2010,Abdelhaye:2012}.
Our observations are summarized in a stability diagram, where we report the radial velocity $v_r=v_l(R/2h_0)$, which is the velocity for radial air invasion, and the initial gap thickness $h_0$ [Fig.~\ref{fig1}(e)]. The transition from unstable to stable displacement occurs at a critical radial velocity $v_r^*$, which increases linearly with increasing $h_0$. Such a scaling is robustly observed for different concentrations of carbon black ranging from 4\% to 10\% wt.~[Fig.~S2 in Supplemental Material]. Remarkably, the stability boundary $v_r^*(h_0)$ shifts towards lower values of velocity for carbon black gels with increasing particle concentration; stronger gels exhibit unstable patterns over a larger range of $v_r$ and $h_0$. From our additional experiments on gels with 4\%, 6\% and 10\% wt. we find $v_r^*/h_0 \propto \sigma_c^{-1}$, as shown in the inset of Fig.~\ref{fig1}(e). This scaling suggests a criterion for fingering instabilities governed by a critical energy provided to the material during the tensile test. Indeed, integrating Darcy's law to compute the pressure field between the plates, which is in turn integrated twice over the area occupied by the material yields the expression for the energy input: $E(h)\propto v_r \sigma_c R^3 (h_0^3/h^4)$ for plates separated by a distance $h$ [see Supplemental Material]. Assuming that the flow is unstable above a critical energy $E^*$ at the start of the tensile test leads to a critical radial velocity $v_r^* \propto (E^*/\sigma_c) (h_0/R^3)$, consistent with our experimental scaling.\\  
\indent This energy criterion is confirmed by our measured input energies calculated from the normal force recorded during the tensile test. Unstable patterns form above a critical energy $E^*$, which increases as a power-law with $h_0$, as shown in Fig.~\ref{fig2}. The $h_0$ dependence accounts for the fractal nature of the gel, which implies that the number of particles within the gap $h_0$ scales as $(h_0/a)^{d_f/3}$ (red line in Fig.~\ref{fig2}). The criterion for fingering instabilities is thus a critical local energy input per carbon black particle, which is independent of the yield stress of the gel and the concentration of particles, as shown in the inset of Fig.~2. Note that $E^*$ is equivalent to a characteristic stress $\sigma^\star \simeq 10$~Pa, a value comparable with the activation stress identified in delayed yielding experiments corresponding to the mean elastic barrier that needs to be overcome for a gel strand to break \cite{Sprakel:2011,Lindstrom:2012,Gibaud:2016}.

\begin{figure*}[!ht]
\centering
\includegraphics[width=0.8\linewidth]{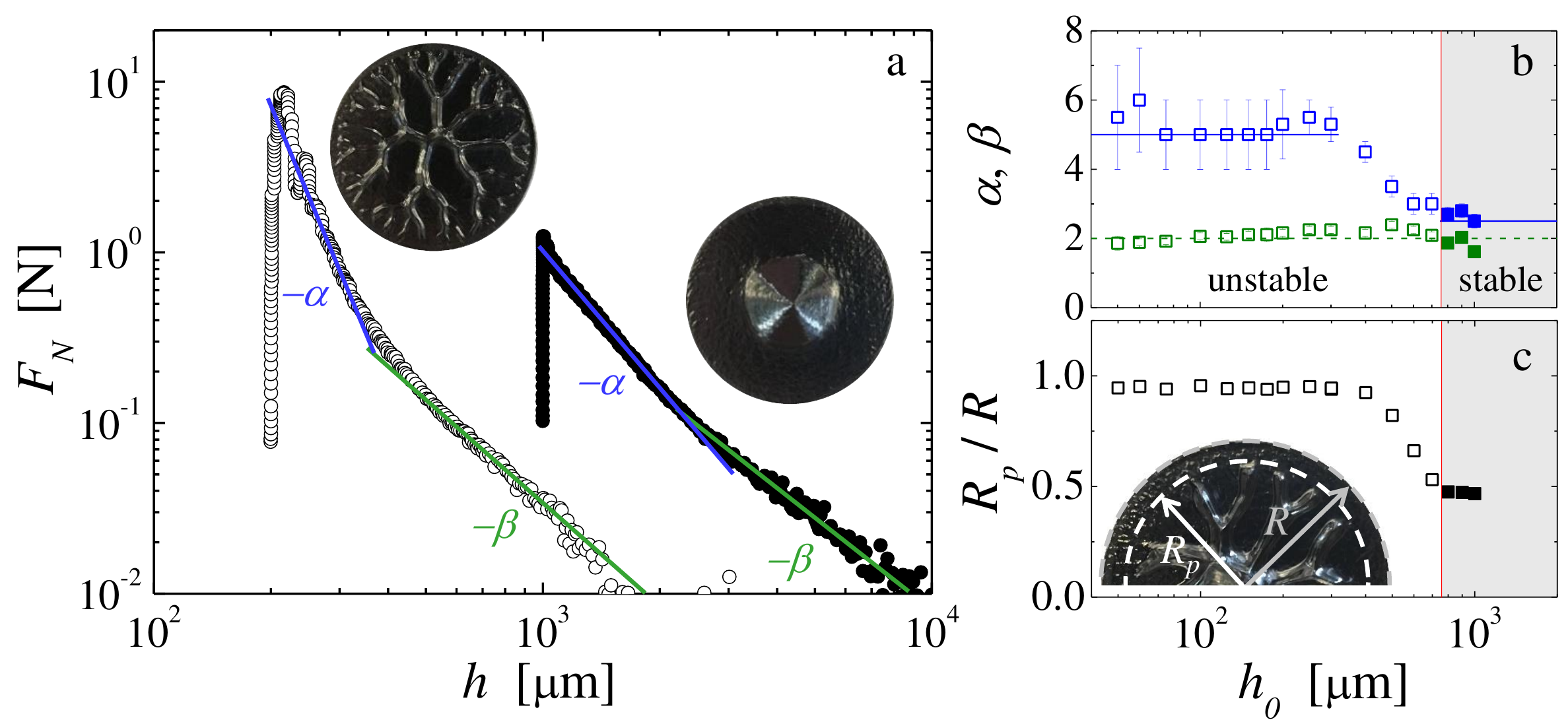}
\caption{(a) Normal force $F_N$ versus gap thickness $h$ during a tensile test ($v_l=200~\mu$m). The two curves correspond to $h_0=200~\mu$m resulting in viscous fingering (open symbols) and $h_0=1000~\mu$m resulting in a stable conical deposit (closed symbols). The blue lines denote the power-law exponents of the first relaxation step: $\alpha=5$ for $h_0=200$~$\mu$m and $\alpha=2.5$ for $h_0=1000$~$\mu$m. The green lines denote the power-law exponents of the second relaxation step: $\beta\simeq 2$ for both experiments. (b) Exponents $\alpha$ and $\beta$ versus $h_0$. With increasing gap thickness, $\alpha$ transitions from a Newtonian-like response ($\alpha=5$ -- blue line) to a yield stress dominated behavior ($\alpha=2.5$ -- blue line). $\beta$ is roughly constant over the range of gap thickness explored ($\beta \simeq 2$ -- green line). (c) The relative extent $R_p/R$ of the pattern versus $h_0$. Inset: definition of $R$ and $R_p$. The vertical lines in (b) and (c) denote the boundary between the stable and the unstable regime.
\label{fig3}}
\end{figure*} 

This scenario suggests that the gel fully yields in the unstable regime, but not in the stable regime. This is further evidenced by considering the evolution of the normal force $F_{N}(h)$ during plate separation for a stable and an unstable pattern. 
In both cases, $F_{N}$ exhibits a sharp increase up to a maximum followed by a two-step relaxation characterized by two power laws with respective exponents $\alpha$ and $\beta$, as shown in Fig.~\ref{fig3}(a). For the unstable pattern, the first relaxation exhibits an exponent $\alpha \simeq 5$ that is characteristic of a purely Newtonian response \cite{Bikerman:1947,Maugis:1991,Lindner:2005}. Indeed, assuming a constant viscosity $\eta$ yields $F_{N}(h) = 3\pi \eta R^4 h_0^2 v_l/2h^5$ \cite{Bikerman:1947,Derks:2003}. The air fingers invade the gap radially into the locally fluidized gel. The second relaxation exhibits a power-law exponent $\beta \simeq 2$, which is characteristic of the necking of a viscous liquid \cite{Bikerman:1947} and corresponds to the extensional flow of the gel threads linking the crest of the branched pattern formed on the upper and lower plates.\\ 
\indent By contrast, the early growth of the stable pattern is characterized by a power-law relaxation step with an exponent $\alpha \simeq 2.5$, characteristic of a yield stress fluid \cite{Coussot:1999b,Derks:2003}. Indeed, assuming that the pressure gradient is balanced by the yield stress $\sigma_c$ yields $F_{N}(h) = 2\pi R^3 h_0^{3/2} \sigma_c/3h^{5/2}$. While being dragged towards the center of the plates, the carbon black gel thus behaves predominantly as an elastic soft solid. The gel may rearrange to accommodate the extensional flow, but over timescales that are large compared to the gel's ``healing timescale'' denoting the re-formation of the network, such that $G'>G''$ at all times. The second relaxation step exhibits a power law with an exponent $\beta \simeq 2$, as for the unstable case. Note that the transition from the first to the second relaxation regime occurs at a critical gap $h_c \simeq 2600~\mu$m that coincides with reaching the final diameter of the deposit, as determined from mass conservation arguments, further confirming that the second relaxation step denotes the thinning of the gel thread connecting the two cones of gel located on the lower and upper plates.\\
\indent More generally, for tensile tests performed at a constant lift velocity, the first relaxation step of the normal force displays a Newtonian behavior characterized by $\alpha\simeq 5$ at small $h_0$. For increasing initial gap thicknesses approaching the boundary to the stable regime, $\alpha$ decreases monotonically until reaching $\alpha\simeq 2.5$, the value expected for a yield stress fluid, at the onset of stable displacement, as shown in Fig.~\ref{fig3}(b). Concomitantly, the pattern transitions from a highly branched structure that extends over the entire radius $R$ of the plate, to an unstable pattern of reduced size $R_p$ within a transitional range of $h_0$ to finally a stable conical shape, as shown in  Fig.~\ref{fig3}(c). 
The second relaxation step of the normal force displays a power-law exponent $\beta \simeq 2$ for all initial gap thicknesses, whether or not the flow is unstable. These observations are robust and also observed in experiments at fixed $h_0$ performed at various lift velocities [Fig.~S3 in Supplemental Material].

\begin{figure}[!b]
\centering
\includegraphics[width=0.9\linewidth]{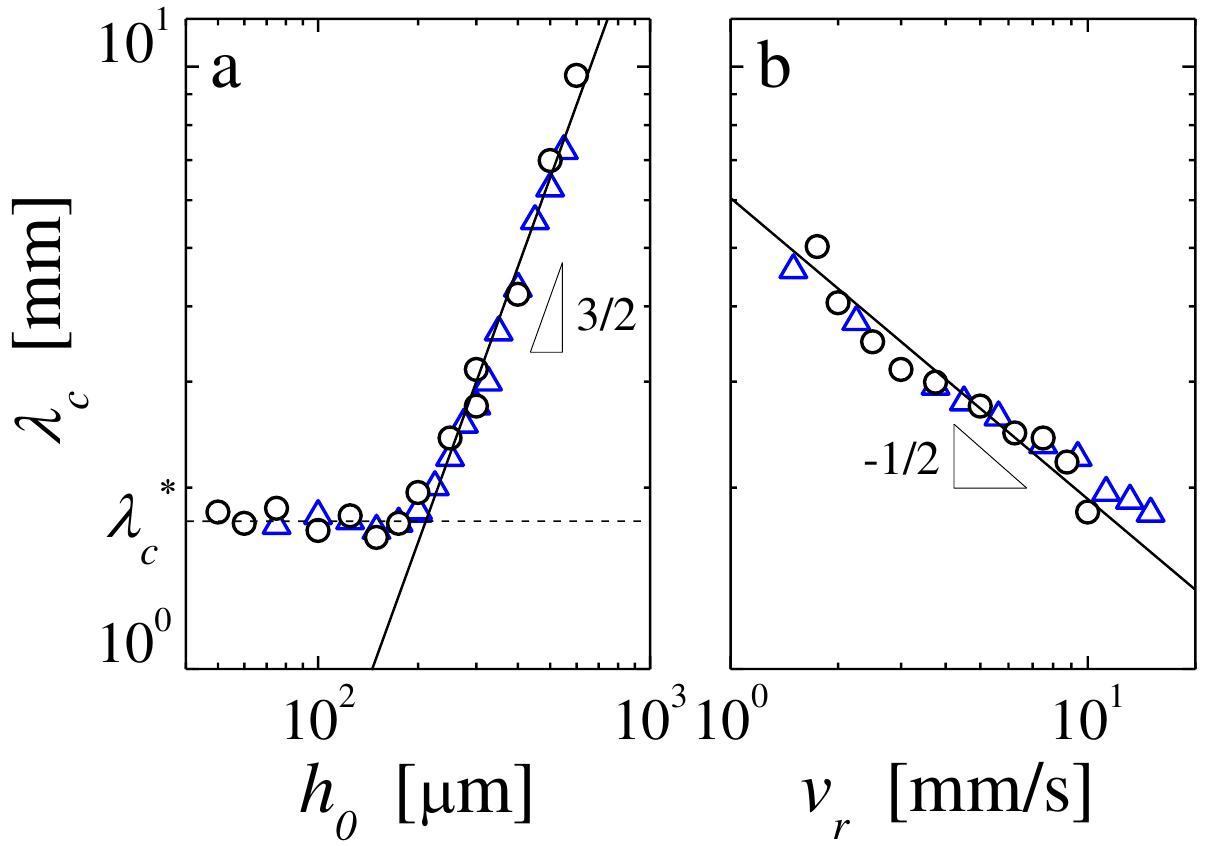}
\caption{Most unstable wavelength $\lambda_c$ of the fingering patterns versus (a) the initial gap thickness $h_0$ and (b) the radial velocity $v_r$. In (a), the solid line corresponds to a power-law exponent of $3/2$, the dashed line to $\lambda_c^*=1.76$~mm. In (b), the line corresponds to a power-law exponent of $-1/2$. Data obtained with a 8\% wt.~carbon black gel and two different plate radii $R=20$~mm ($\circ$) and $30$~mm (\textcolor{blue}{$\triangle$}).
\label{fig4}}
\end{figure}

As a consequence of the gel's full fluidization locally at the tip of the fingers, the most unstable wavelength characterizing the onset of the instability $\lambda_c$ \cite{CBfootnote1} scales as $\lambda_c \propto h_0^{3/2}$ with the initial gap thickness for $h_0 > 200~\mu$m, and as $\lambda_c \propto v_r^{-1/2}$ with the radial velocity, as shown in Fig.~\ref{fig4}. These two scaling laws are in agreement with the predictions from a linear stability analysis for a Newtonian fluid with viscosity $\eta$ and surface tension $\Gamma$ \cite{Paterson:1981,Shelley:1997}: 
\begin{equation}\lambda_c \simeq \sqrt{\frac{2\pi h_0^3 \Gamma}{\eta v_lR}} = \frac{\pi h_0}{\sqrt{\eta v_r/\Gamma}}
\end{equation}
for plates with large aspect ratio $R/h_0 \gg 1$. This confirms that the yield stress plays a negligible role in the formation of the fingering patterns. 
Moreover, if we take for $\Gamma$ the surface tension of the light mineral oil $\Gamma_s$, we obtain $\eta=0.18$~Pa.s, which is compatible with the viscosity of a fully fluidized carbon black gel measured at high shear rate $\dot \gamma=1000$~s$^{-1}$ \cite{Grenard:2014,CBfootnote2}.
Remarkably, the most unstable wavelength saturates at a constant value $\lambda_c^*=1.76\pm$0.10~mm for $h_0<200~\mu$m, as shown in Fig.~\ref{fig4}(a). Such saturation is unexpected. We note, however, that this value is compatible with the capillary length $\ell_c = \sqrt{\Gamma_s/(\rho_s g)} \simeq1.8$~mm of the mineral oil, which suggests that the thinnest fingers that form during a tensile experiment are limited by capillary effects.

In summary, we show that the presence of attractive colloidal particles suspended in a Newtonian liquid can suppress the viscous fingering instability. We provide the criterion for the onset of the instability as a critical energy input needed to break bonds between particles allowing for the local fluidization of the gel and the invasion of fingers. These air fingers invade the gap producing highly branched patterns that exhibit wavelengths and normal-force responses characteristic of Newtonian fluids.
This scenario should apply to clays and other attractive suspensions, where the normal-force response associated with unstable patterns is reported to be viscous dominated \cite{Abdelhaye:2008}. Moreover, the observation that the fingering characteristics are set solely by the properties of the fully fluidized state of the sample might shed light on discrepancies pointed out in \cite{Derks:2003}, where the finger width observed in hair gel solutions is independent of their yield stress.
More generally, the local fluidization scenario described here is in line with recent rheological work on stress-induced failure in gels \cite{vanDoorn:2018}, suggesting that our critical energy criterion could be relevant for predicting the outcome of delayed failure in colloidal gels \cite{Cipelletti:2020,Gibaud:2020}.
However, our results strongly contrast with experiments on jammed assemblies of soft particles such as dense microgels for which the wavelength of the pattern is set by the yield stress of the material \cite{Lindner:2000}. Indeed, the mechanism of viscous fingering instabilities described here is unique to yield stress fluids composed of attractive particles at low volume fractions, and we expect different scenarios to govern unstable growth in soft repulsive glasses; our work indicates that one should not look for a universal scenario describing viscous fingering in all yield stress fluids.

\begin{acknowledgments}
We thank S. Manneville, J.A. Dijksman, A. Helal, and G.H. McKinley for fruitful discussions. T.D. acknowledges support from the National Science Foundation under Grant No. NSF PHY 17-48958 through the KITP program on the Physics of Dense Suspensions. B.M. and I.B. acknowledge support from MIT MISTI-France.
\end{acknowledgments}


\begin{thebibliography}{57}
\bibitem [{CBf({\natexlab{a}})}]{CBfootnote1}%
  \BibitemOpen
  \href@noop {} {} \bibinfo {note} {The most unstable
  wavelength $\lambda_{\rm c}$ is determined by counting the number of fingers
  along the periphery of the plate. The underlying assumption that the pattern
  at the plate periphery is representative of the early stage of the experiment
  is justified as the short healing time of the gel, and the gel's solid-like
  behavior at rest, allow for the early-stage pattern to be frozen in and to
  remain unchanged from the time of formation. Note that a direct visualization
  of the early stage is not possible because carbon black gels are optically
  opaque and leave a thin layer of gel adhered to the plates.}\BibitemShut
  {Stop}%
\bibitem [{CBf({\natexlab{b}})}]{CBfootnote2}%
  \BibitemOpen
  \href@noop {} {} \bibinfo {note} {This viscosity is about
  a factor of 4 lower than the one extracted from the first relaxation step of
  the normal force. This is because the viscosity extracted from $\lambda_c$
  corresponds to the state of the gel at the very tip of the finger, whereas
  that extracted from the force relaxation is an average all along the length
  of the fingers.}\BibitemShut {Stop}%
\end{thebibliography}

%


\clearpage
\newpage
\onecolumngrid
\setcounter{page}{1}

\begin{center}
    {\large\bf Criterion for fingering instabilities in colloidal gels}
\end{center}
\begin{center}
    {\large\bf Supplemental Material}
\end{center}
\vspace{-0.4cm}
\setcounter{figure}{0}
\global\def\thefigure{S\arabic{figure}}
\setcounter{table}{0}
\global\def\thetable{S\arabic{table}}
\section{Energy input associated with the plate separation}
\vspace{-0.2cm}
At the onset of the instability the gel is fluidized locally at the locus of the growing fingers. The pressure gradient in the gel reads ${\rm d}p /{\rm d} r= [1+ f(\dot \gamma)]\sigma_c/h$, where $\sigma_c$ denotes the yield stress of the gel, $h$ the gap spacing, and $f$ represents a function such that $f(\dot \gamma)\rightarrow 0$ in the limit of vanishing shear rate \cite{Coussot:1999b}. In the limit of high shear rate, the rheology of carbon black gels is dominated by the viscous term, such that ${\rm d}p /{\rm d} r \simeq f(\dot \gamma)\sigma_c/h$, with $f(\dot \gamma) \propto v_r/h$, since carbon black dispersions behave as a Bingham fluids in the limit of high shear rate \cite{Trappe:2000}. Substituting the radial velocity with the lift velocity, $v_r=v_l\,r/h$, and using volume conservation, $hr^2=h_0R^2$, we can integrate the pressure gradient to find $p(r)$, which in turn can be integrated to determine the normal force $F_N(h)=\int 2\pi r p(r){\rm d}r$. The energy input is then $E(h) \propto \int F_N(h)dh \propto v_r\sigma_cR^3(h_0^3/h^4)$.
\section{Linear viscoelastic properties of carbon black gels}
\vspace{-0.5cm}
\begin{figure*}[!h]
\centering
\includegraphics[width=0.8\linewidth]{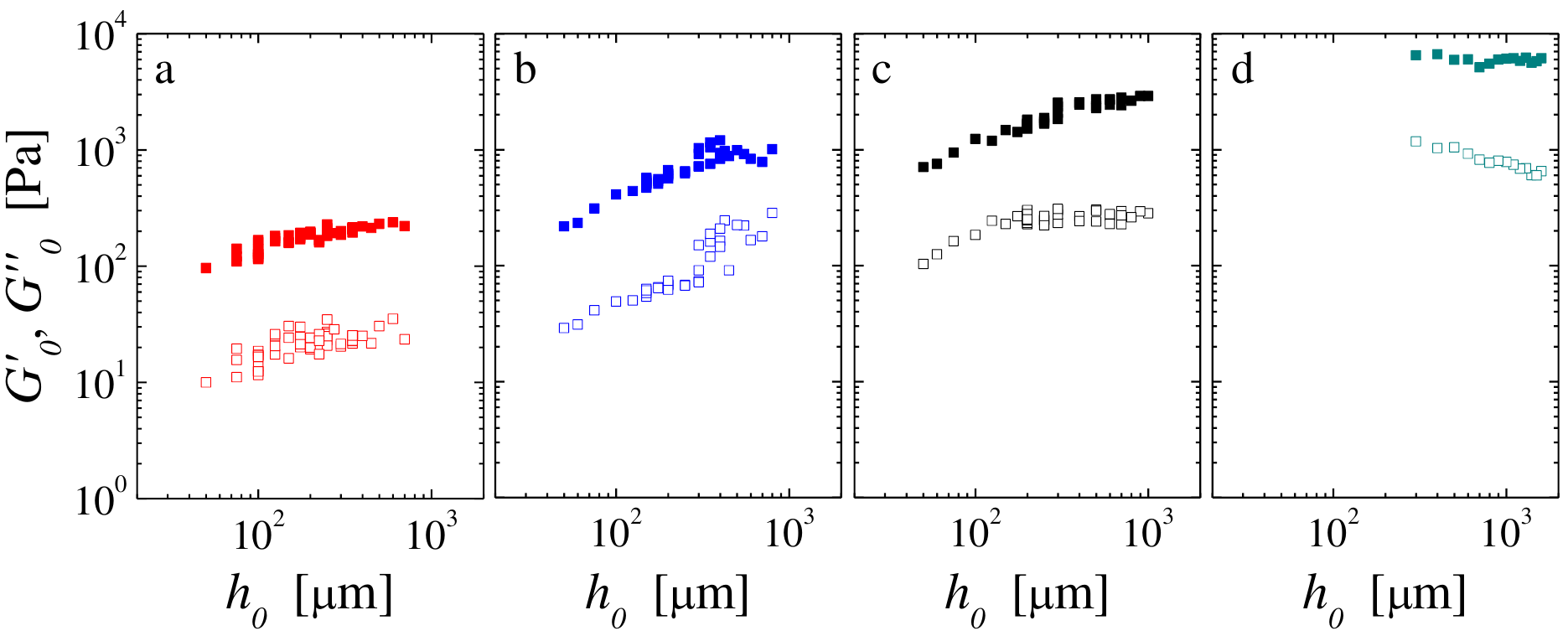}
\caption{Linear viscoelastic moduli $G'_0$ ($\blacksquare$) and $G''_0$ ($\square$) versus the initial gap thickness $h_0$ for carbon black gels of concentrations: 4\%, 6\%, 8\% and 10\% wt. from (a) to (d). Measurements are performed under an imposed shear strain $\gamma_0=$0.1\% at a frequency $f=1$~Hz. Prior to each measurement, the sample is fully fluidized by imposing a large preshear at a stress $\sigma=100$~Pa for 30~s. The gel reforms during a stress sweep from 100~Pa down to 0~Pa at constant rate of 1~Pa.s$^{-1}$. The linear viscoelastic moduli are measured for 60~s and we report the average values over the last 30~s.
\label{figS1}}
\end{figure*} 
\vspace{-0.5cm}
\section{Stability diagram for gels of different concentrations}
\vspace{-0.4cm}
\begin{figure*}[!h]
\centering
\includegraphics[width=0.8\linewidth]{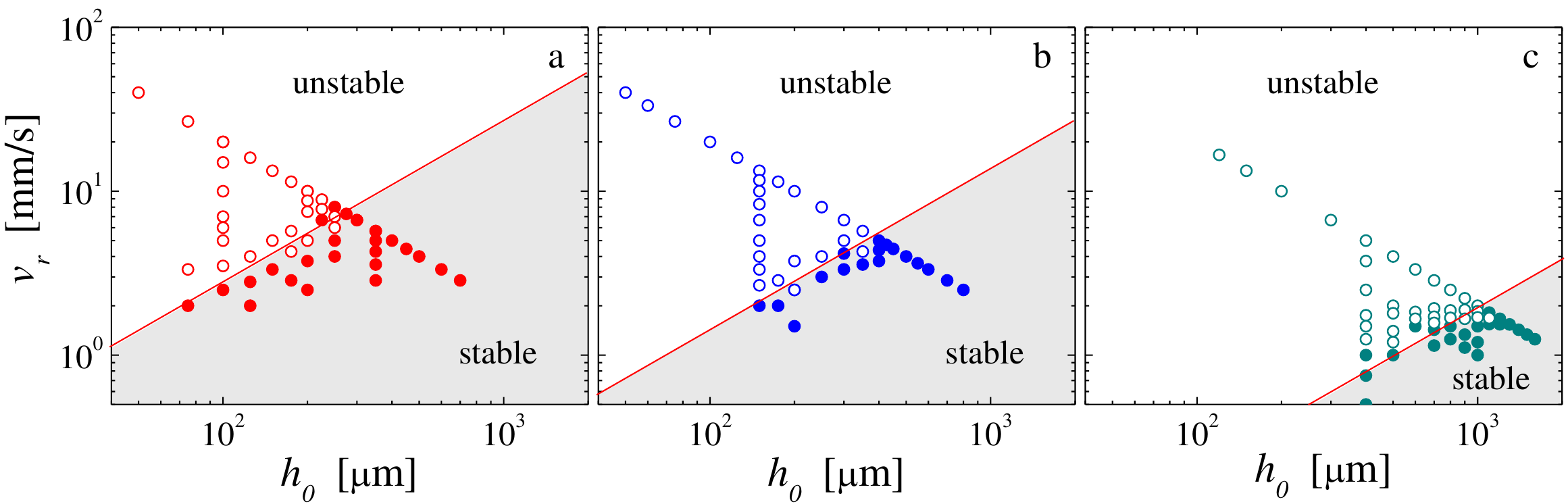}
\caption{Stability diagram reporting $v_r$ versus $h_0$ for tensile tests performed with carbon black gels of concentrations 4\%, 6\% and 10\% wt.~from (a) to (c). The red line denotes the critical velocity $v_r^*(h_0)$ separating the stable from the unstable regime. The boundary between stable and unstable flow shifts towards lower velocities with increasing carbon black concentration.
\label{figS2}}
\end{figure*} 
\newpage
\section{Normal force relaxation at different lift velocities}
\vspace{-0.4cm}
\begin{figure}[!h]
\centering
\includegraphics[width=0.8\linewidth]{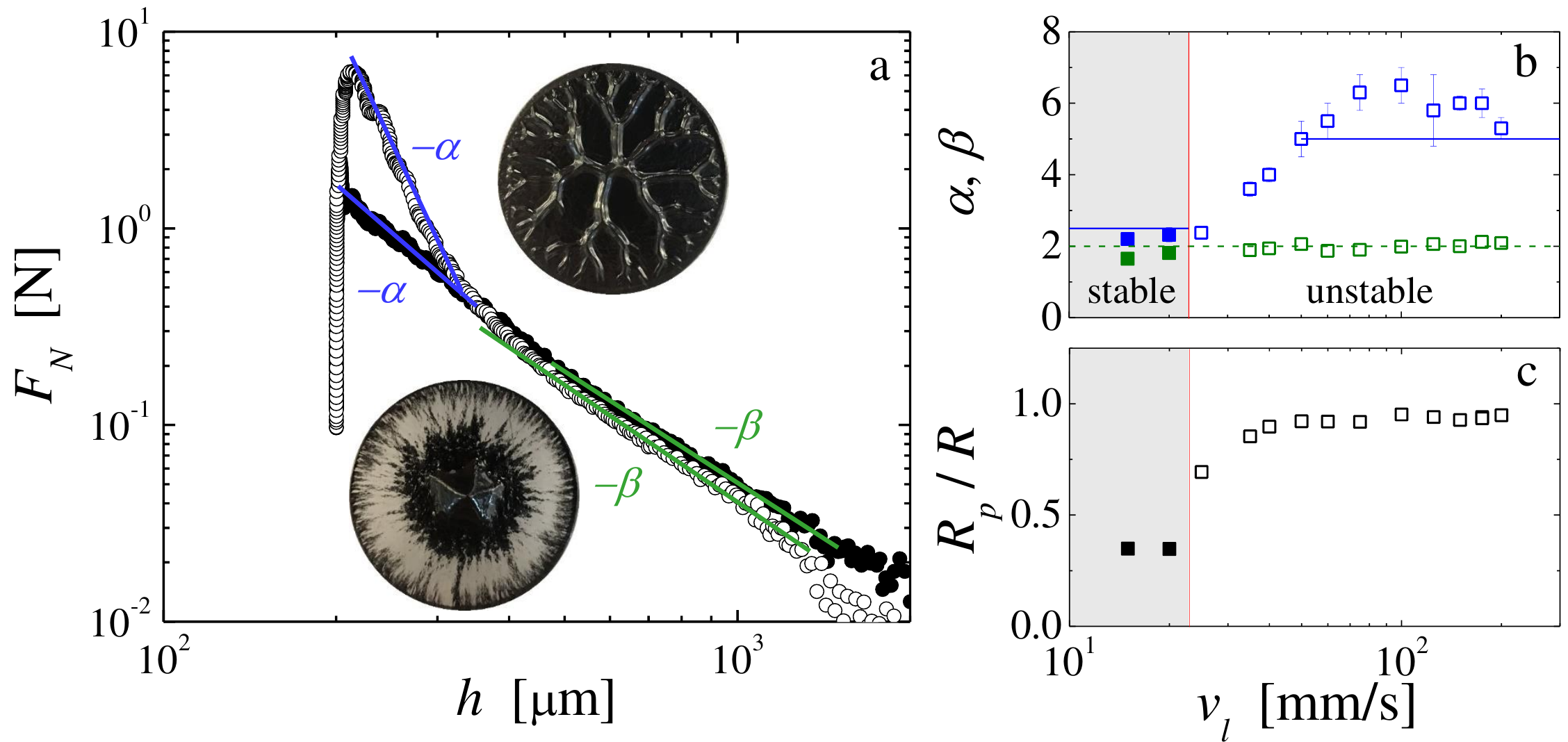}
\caption{(a) Normal force $F_{N}$ versus gap thickness $h$ during a tensile test performed at lift velocities $v_l=20$~$\mu$m/s (closed symbols) and 100~$\mu$m/s (open symbols) in a parallel plate geometry of radius $R=20$~mm. The blue lines denote the power-law exponents of the first relaxation step: $\alpha\simeq 2.5$ for $v_l=20$~$\mu$m/s and $\alpha\simeq 5$ for $v_l=100$~$\mu$m/s. The green lines denote the power-law exponents of the second relaxation step: $\beta\simeq 2$ for both experiments. (b) Exponents $\alpha$ and $\beta$ versus the lift velocity $v_l$. With increasing lift velocity, $\alpha$ transitions from a yield stress dominated behavior ($\alpha=2.5$ -- blue line) to a Newtonian-like response ($\alpha=5$ -- blue line). $\beta$ is roughly constant over the range of lift velocities explored ($\beta \simeq 2$ -- green line). (c) The relative extent $R_p/R$ of the pattern versus $v_l$. The vertical lines in (b) and (c) denote the boundary between the stable and the unstable regime.
\label{figS3}}
\end{figure} 

\end{document}